\documentclass[aps,pra,preprint,groupedaddress,amsmath,amssymb,preprintnumbers,showpacs]{revtex4-1}

\usepackage{accents}
\usepackage{enumerate}
\usepackage{graphicx}
\usepackage{color}
\usepackage{float}
\graphicspath{{./}}

\newcommand{\ket}[1]{\bigl| #1 \bigr>} 
\newcommand{\bra}[1]{\bigl< #1 \bigr|} 
\newcommand{\braket}[2]{\bigl< #1 \vphantom{#2} \bigr|
 \bigl. #2 \vphantom{#1} \bigr>} 

\newcommand{\abs}[1]{\left| #1 \right|} 
\newcommand{\comm}[2]{\Bigl[#1,#2\Bigr]}
\renewcommand{\d}[2]{\frac{\text{d} #1}{\text{d} #2}} 
\newcommand{\pd}[2]{\frac{\partial #1}{\partial #2}} 
\newcommand{\pdd}[2]{\frac{\partial^2 #1}{\partial #2^2}} 
\newcommand{\pdt}[3]{\frac{\partial^2 #1}{\partial #2 \partial #3}}
\newcommand{\at}[2][]{#1|_{#2}}

\begin{document}

\title{Dissociation and dissociative ionization of ${\text{H}_2}^+$ using the time-dependent surface flux method}

\author{Lun \surname{Yue}}
\author{Lars Bojer \surname{Madsen}}
\affiliation{Department of Physics and Astronomy, Aarhus University, DK-8000 Aarhus C, Denmark}
\date{\today}

\begin{abstract}
The time-dependent surface flux method developed for the description of electronic spectra [L. Tao 
and A. Scrinzi, New J. Phys. \textbf{14}, 013021 (2012); A. Scrinzi, New J. Phys. \textbf{14}, 085008 
(2012)] is extended to treat dissociation and 
dissociative ionization processes of ${\text{H}_2}^+$ interacting with strong laser pulses. 
By dividing the simulation volume into proper spatial regions associated with the individual reaction 
channels and monitoring 
the probability flux, the joint energy spectrum for the dissociative ionization process and the energy 
spectrum for dissociation is obtained. The methodology is illustrated by solving the time-dependent 
Schr\"{o}dinger equation (TDSE) for a 
collinear one-dimensional model of ${\text{H}_2}^+$ with electronic and nuclear motions treated exactly and 
validated by 
comparison with published results for dissociative ionization.  The results 
for dissociation are qualitatively explained by 
analysis based on dressed diabatic Floquet potential energy curves, and the method is used to 
investigate the 
breakdown of the two-surface model.
\end{abstract}

\pacs{33.80.Rv, 31.15.xv, 33.20.Xx}

\maketitle

\section{Introduction}
The advent of new extreme ultraviolet sources with femtosecond or subfemtosecond duration 
\cite{Sansone11,MacNeil10} has sparked interest in measuring the dynamics of atoms and molecules on their natural time scales 
\cite{Krausz09}. Typically, to obtain time-resolved information, pump-probe schemes have been used, where a pump 
pulse induces dynamics in the system under investigation and a subsequent probe pulse is applied after a fixed time delay to 
extract time information. Realizations of such pump-probe methodologies in the subfemtosecond time regime include 
attosecond streaking spectroscopy \cite{Kienberger02,Itatani02,Schultze10} and attosecond interferometry 
\cite{Remetter06,Klunder11}, and these schemes involve couplings  to at least a single continuum, which from a theoretical and 
computational standpoint is challenging due to the requirement of an accurate description of the continua. 
Extraction of the relevant observables such as the correlated momentum or energy distributions of the 
final fragments introduces another numerical obstacle. One method is to wait until the bound and scattered parts of the 
wave function are separated, whereafter a projection of the scattered part onto asymptotic channel eigenstates is 
performed \cite{Laulan04,Feist08,Madsen07}. This puts a lower limit on the size of the simulation volumes as the 
scattered part of the wave function must not reach the volume boundaries. Depending on field parameters, this usually 
requires a huge simulation volume, which is computationally challenging. Another method is to project the total wave function onto 
exact scattering states at the end of the time-dependent pertubation \cite{Fernandez09}. The advantage of this is that a 
smaller simulation volume can be used compared to the former method, while the drawback is the difficulty in obtaining 
the exact scattering states for a single continuum and the nonexistence of exact scattering states for several continua.

Recently, the time-dependent surface flux (t-SURFF) method \cite{Tao12,Scrinzi12} was introduced to 
extract fully differential ionization spectra of one- and two-electron atomic systems from numerical time-dependent Schr\"{o}dinger equation (TDSE)
TDSE calculations.
By placing absorbers at the grid boundaries that absorbed the outgoing electron flux, the total number 
of discretization points used was decreased, thereby reducing the numerical effort. Energy spectra are 
still obtainable by monitoring the flux passing through surfaces placed at distances smaller than 
the absorber regions.
Several related flux methods already exist, e.g., the ``virtual detector" method \cite{Feuerstein03a} and 
the wave-function splitting technique \cite{Keller95,Chelkowski98}. 
The advantage of the t-SURFF method is that complete information in a given reaction channel can be 
obtained.

The aim of this work is to demonstrate that the t-SURFF method can be extended to treat molecular 
systems interacting with a laser field, including both the electronic and nuclear degrees of freedom.
The simplest molecule, ${\text{H}_2}^+$, is of particular interest to physicists, since understanding the 
fundamental physical processes in this molecule will add valuable insight into more complex molecular 
systems. There has been impressive experimental and theoretical progress over the past decades in 
the understanding of laser-induced dissociation and dissociative ionization (DI) phenomena in H$_2^+
$ \cite{Posthumus04,Giusti-Suzor95}. The phenomena described include charge-resonance enhanced 
ionization \cite{Zuo95}, bond softening \cite{Bucksbaum90}, bond hardening 
\cite{Yao92,Yao93}, above-threshold dissociation (ATD) \cite{Jolicard92,Giusti-Suzor90}, high-order-harmonic generation \cite{Zuo93},  and above-threshold Coulomb explosion
\cite{Esry06}.

Much of the insight in these processes comes from calculations involving exact numerical propagation 
of the TDSE for  ${\text{H}_2}^+$ model systems with reduced 
dimensionality 
\cite{Kulander96,Chelkowski98,Feuerstein03b,Takemoto10,Madsen12,Silva13,Steeg03,Weixing01}. 
For ${\text{H}_2}^+$ interacting with a laser pulse, one of the essential questions is how the energy of the 
pulse is distributed between the resulting fragments. For the DI process, the desired observable is the 
joint energy spectrum (JES), which displays the differential probability for simultaneously detecting a 
particular electronic kinetic energy and a particular nuclear kinetic energy \cite{Madsen12}. The JES 
for DI of H$_2$ has been considered theoretically in the single-photon ionization regime~
\cite{Gonzalez12,Silva12} 
and experimentally in the multiphoton regime~\cite{Wu13}. Theoretically, the JES of ${\text{H}_2}^+$ was 
obtained from numerical TDSE calculations by projecting on approximate double continuum 
eigenstates at the end of the pulse \cite{Madsen12}, and by using a resolvent technique \cite{Silva13}. 
In both theoretical works the TDSE was solved on a large numerical grid with many discretization 
points to ensure that the DI wave packet did not reach the grid boundaries during the pulse duration. 
For longer pulses, the increase in the discretization points can make the computational effort 
unmanageable.

The t-SURFF method, when applied to ${\text{H}_2}^+$, circumvents the before-mentioned deficiencies, 
and can be used to obtain complete information in a given reaction channel.
Hence, not only is it possible to obtain the total nuclear kinetic energy release (KER) spectra for 
dissociation and DI, the JES for DI and the KER spectrum in each electronic dissociation channel can 
be obtained as well. Furthermore, the t-SURFF method is extendable to systems with more degrees of 
freedom, due to the usage of small simulation volumes, and circumvents also in this case 
the construction of scattering states to extract observables.
This makes the t-SURFF method attractive for the study of the laser-induced  time-dependent 
laser-molecule interaction. 

The paper is organized as follows. In Sec.~\ref{sec:model}, the ${\text{H}_2}^+$ model system and the 
electromagnetic field is described. In Sec.~\ref{sec:theory}, the t-SURFF method for ${\text{H}_2}^+$ is 
explained. In Sec.~\ref{sec:results}, the kinetic energy spectra for dissociation and DI from TDSE 
calculations are obtained for different field parameters. For the DI process, the JES are obtained, 
while for the dissociation process, the KER for 
dissociation into H($n=1$) and H($n=2$)
are obtained. 
The concluding remarks are contained in Sec.~\ref{sec:conclusion}. 
Atomic units are used throughout, unless indicated otherwise.

\section{Model for ${\text{H}_2}^+$}\label{sec:model}

To reduce the numerical effort, a simplified model with reduced dimensionality of ${\text{H}_2}^+$ is employed 
that includes only the dimension that is aligned with the linearly polarized laser pulse. Within this 
model, electronic and nuclear degrees of freedom are treated exactly. Such models have been used 
extensively in the literature and reproduce experimental results at least qualitatively 
\cite{Kulander96,Steeg03,Weixing01}. 

After separating out the center of mass motion of the nuclei, the TDSE for the model ${\text{H}_2}^+$ 
molecule in the dipole approximation and velocity gauge reads
\begin{equation}
  i\partial_t\ket{\Psi(t)}=H(t) \ket{\Psi(t)}
  \label{eq:TDSE}
\end{equation}
with the Hamiltonian
\begin{equation}
  \begin{aligned}
    H(t)
    =T_\text{e}+T_\text{N}+V_\text{eN}+V_\text{N}+V_\text{I}(t),
    \label{eq:Hamiltonian}
    \end{aligned}
\end{equation}
where  $\ket{\Psi(t)}$ in position space depends on the internuclear distance $R$ and the electronic 
coordinate  $x$  measured with respect to the center-of-mass of the nuclei. The components of the 
Hamiltonian in Eq.~\eqref{eq:Hamiltonian} are $T_\text{e}=-(1/2\mu)\partial^2/\partial x^2$, $T_\text{N}
=-(1/m_\text{p})\partial^2/\partial R^2$, $V_\text{eN}=-1/\sqrt{{(x-R/2)^2+a(R)}}-1/\sqrt{(x+R/2)^2+a(R)}
$, $V_\text{N}=1/R$, and $V_\text{I}(t)=-i \beta A(t) \partial/\partial x $,
where $m_\text{p}=1.836\times 10^3$ a.u. is the proton mass, $\mu=2m_\text{p}/(2m_\text{p}+1)$ is 
the reduced electron mass, $\beta = (m_\text{p}+1)/m_\text{p}$, and the softening parameter $a(R)$ for 
the Coulomb singularity is chosen to produce the exact three-dimensional $1s\sigma_g$ Born-Oppenheimer (BO) potential energy curve \cite{Madsen12}.

The vector potential that we use is of the form
\begin{equation}
  A(t)=A_0 \sin^2\left(\frac{\pi t}{T_\text{pulse}}\right)\cos(\omega t),
  \label{eq:vecpot}
\end{equation}
with  the angular frequency $\omega$, and the pulse duration  $T_\text{pulse}$ related to the number 
of optical cycles $N_\text{c}$ by $T_\text{pulse}=N_\text{c}2\pi/\omega$. The amplitude $A_0$ is 
chosen such that $\omega^2A_0^2=I$, with $I$ the intensity. 
In this work, we consider laser intensities in the range $I=10^{13}$-$10^{14}$ W/cm$^2$, laser 
frequencies of $400$ and $800$ nm, and the number of optical cycles $N_\text{c}=10$.

Equation \eqref{eq:TDSE} is solved exactly on a two-dimensional spatial grid using the split-operator, 
fast Fourier transform (FFT) method \cite{Feit82}, with a time step of $\Delta t=0.005$ in the time propagation. The grid size is defined by $\abs{x}\le 100$ and  $R
\le 40$, with grid spacings $\Delta x=0.781$ and $\Delta R=0.078$. Complex absorbing potentials 
(CAPs) are used to absorb the outgoing flux and to avoid reflections at the grid boundaries. Indeed, it 
is the introduction of the CAP that allows us to use a grid size that is numerically manageable. The 
form of the CAP is \cite{muga04}
\begin{equation}
  V_\text{CAP}(r)=
  \begin{cases}
  -i\eta \left(\abs{r}-r_\text{CAP}\right)^n  &, |r| \geq r_\text{CAP} \\
  0 &, \text{elsewhere}  
  \end{cases}
  \label{eq:CAP}
\end{equation}
with $r$ being either the electronic coordinate $x$ or the nuclear coordinate $R$. We use  $
\eta_e=0.001$, $x_\text{CAP}=55$, and $n_e=2$ for the electronic CAP, and $\eta_N=0.01$, $R_
\text{CAP}=26$ and $n_N=2$ for the nuclear CAP.

The size of the simulation volume used should be compared with the sizes of the volumes used by other methods.  Two other methods have been used to determine the JES based on wave-packet propagation. In one work \cite{Madsen12}, a projection on approximate scattering states was performed and a grid with $\abs{x} \leq 1500$ was used for the electronic coordinate. In another work \cite{Silva13}, a resolvent technique was used and a grid with $\abs{x} \leq 3000$ was used for the electronic coordinate. Both these values significantly exceed the size of $\abs{x} \leq 100$ used here.

\section{$\textbf{t}$-SURFF for ${\text{H}_2}^+$}
\label{sec:theory}
The t-SURFF method for DI and dissociation is now outlined for the model ${\text{H}_2}^+$ molecule. 
For ${\text{H}_2}^+$ interacting with a laser pulse, there are two continuum channels: 

(1) The DI channel ${\text{H}_2}^+ \rightarrow p + p + e$, where the final asymptotic state consists of two 
protons and one electron separated by large distances. The observable of interest is the JES that 
shows the energy sharing between the protons and the electron. From the JES, the
electronic above-threshold ionization (ATI) and the nuclear KER spectra are obtained by
integrating out the appropriate degrees of freedom.

(2) The dissociation channel ${\text{H}_2}^+ \rightarrow \text{H} + p$, where  ${\text{H}_2}^+$ dissociates into a proton and a 
hydrogen atom in a given state (channel) . The relevant observables are the channel-specific KER 
spectra that show how the nuclear kinetic energies are shared between the different dissociation 
channels.

\begin{figure}
  \centering
  \includegraphics[scale=0.9]{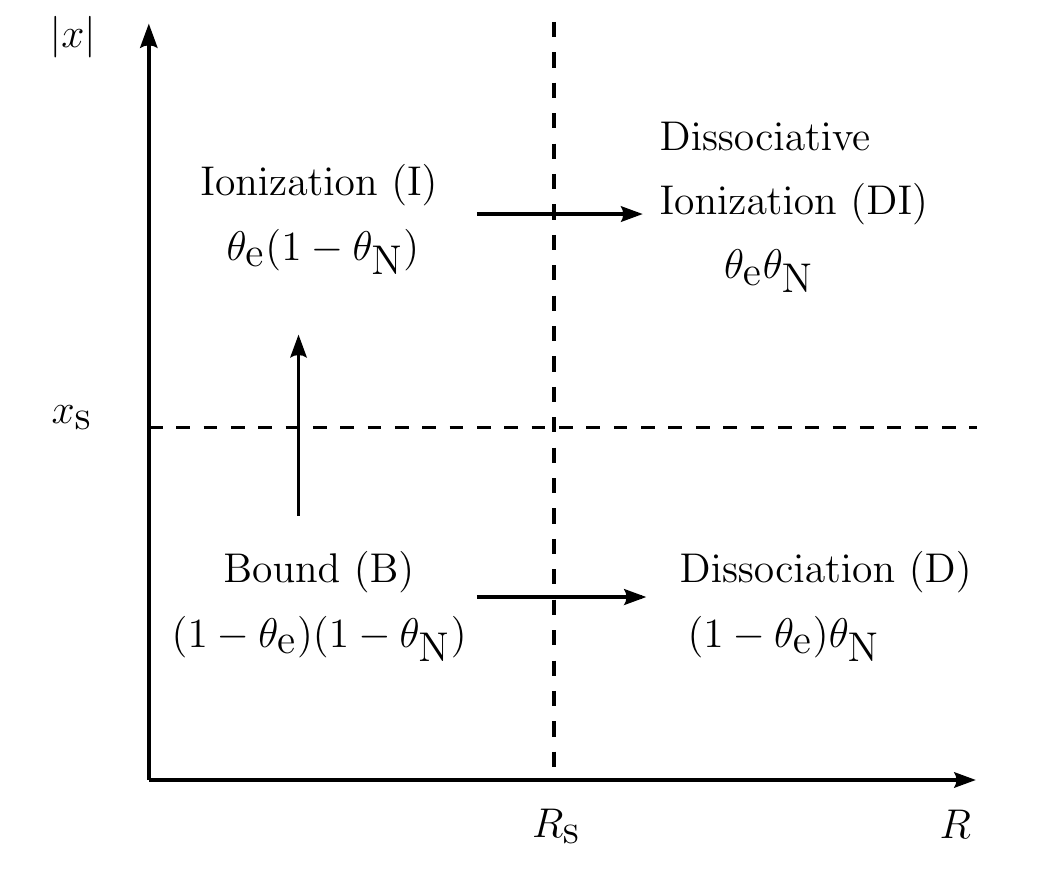}
  \caption{Illustration showing the four spatial regions used to analyze the wave packet
formed by the interaction of ${\text{H}_2}^+$ with the external laser pulse.
The dashed line at $\abs{x}=x_\text{s}$ is a boundary surface beyond which the electron-nuclear interaction 
$V_\text{eN}$ is neglected in the DI channel, while the dashed line at $R=R_\text{s}$ is a boundary 
surface beyond which the nuclear repulsion $V_\text{N}$ is neglected. In the t-SURFF method, the flux 
passing through these surfaces is monitored and used to construct the differential probability 
amplitudes. In the figure, the projection operators, formed from Eqs.~\eqref{eq:proje} and 
\eqref{eq:projN}, that project on the different spatial regions are given in the corresponding reaction 
channels.}
  \label{fig1}
\end{figure}

To identify the different channels and the corresponding observables, we partition the total coordinate 
space into four regions as shown in Fig.~\ref{fig1}. Each region corresponds to a reaction channel. By 
monitoring the flux going through the surfaces at $\abs{x}=x_\text{s}$ and $R=R_\text{s}$, the JES for DI 
and the channel-specific KER for dissociation can be constructed.

To proceed formally, we define the projection operators
\begin{subequations}
  \begin{align}
    \theta_\text{e} &=\int  dx \theta(\abs{x}-x_\textrm{s}) \ket{x} \bra{x}, \label{eq:proje}\\
    \theta_\text{N} &=\int  dR \theta(R-R_\textrm{s}) \ket{R} \bra{R}, \label{eq:projN}
  \end{align}
\end{subequations}
where  $x_\text{s}$ and $R_\text{s}$ are locations of the surfaces beyond which the Coulomb 
interactions $V_\text{eN}$ and $V_\text{N}$ are neglected, respectively, and $\theta(x)$ is the 
Heaviside step function. These projection operators are used to partition the total wave function  into the four parts belonging to the different spatial regions of Fig.~\ref{fig1}:
\begin{equation}
  \begin{aligned}
    \ket{\Psi(t)}
    = \ket{\Psi_\text{B}(t)}+\ket{\Psi_\text{D}(t)}+\ket{\Psi_\text{I}(t)}+\ket{\Psi_\text{DI}(t)},
  \end{aligned}
\end{equation}
with $\ket{\Psi_\text{B}(t)}=(1-\theta_\text{e})(1-\theta_\text{N})\ket{\Psi(t)}$, $\ket{\Psi_\text{D}(t)}=(1-\theta_\text{e})\theta_\text{N}\ket{\Psi(t)}$, $\ket{\Psi_\text{I}(t)}=\theta_\text{e}(1-\theta_\text{N})\ket{\Psi(t)}$, and
$\ket{\Psi_\text{DI}(t)}=\theta_\text{e}\theta_\text{N}\ket{\Psi(t)}$. For sufficiently large times after the 
end of the laser pulse $T>T_\text{pulse}$, the dissociation and DI wave packets will have 
moved into their specific spatial regions such that $\ket{\Psi_\text{B}(T)}$ contains the bound part of 
the total wave packet,  $\ket{\Psi_\text{D}(T)}$ contains the dissociative part, and $\ket{\Psi_\text{DI}
(T)}$ contains the DI part.  At time $T$, the wave packet in the spatial region corresponding to 
ionization $\ket{\Psi_\text{I}(T)}=\theta_\text{e}(1-\theta_\text{N})\ket{\Psi(T)}=0$, 
as all the ionized parts will have moved into the DI region since the nuclei do not support bound states
after the removal of the electron.

\subsection{t-SURFF for dissociative ionization}
With the partitioning of space and wave functions, we are now ready to consider the formulation of the 
t-SURFF methodology for the DI channel  ${\text{H}_2}^+ \rightarrow p + p + e$. The projected TDSE on the 
spatial region describing this channel reads (see Fig.~\ref{fig1})
\begin{equation}
  \begin{aligned}
    i \partial_t\ket{\Psi_\text{DI}(t)}&=H_\text{DI}(t)\ket{\Psi(t)},
    \label{eq:TDSE-DI}
  \end{aligned}
\end{equation}
where we have defined the projected Hamiltonian $H_\text{DI}(t)=\theta_{e}\theta_\text{N}H(t)=T_
\text{e}+T_\text{N}+V_\text{I}(t)$. It is important to notice that $\ket{\Psi(t)}$, not $\ket{\Psi_\text{DI}(t)}
$, appears on the right-hand side of Eq.~\eqref{eq:TDSE-DI}. This reflects that $H(t)$ does not 
commute with $\theta_\text{e}\theta_\text{N}$ for all times $t$.

The TDSE for $H_\text{DI}(t)$ is separable in the electronic and nuclear degrees of freedom, with the 
electronic TDSE given by
\begin{equation}
  \begin{aligned}
      i  \partial_t\ket{\phi(t)}=\left[T_\text{e}+V_\text{I}(t)\right]\ket{\phi(t)},
      \label{eq:TDSE-DI2}
 \end{aligned}
\end{equation}
and the nuclear TDSE given by
\begin{equation}
  \begin{aligned}
      i  \partial_t\ket{\chi(t)}= T_\text{N}\ket{\chi(t)}.
      \label{eq:TDSE-DI3}
  \end{aligned}
\end{equation}
A complete set of the solutions in position space is formed by the Volkov waves $\phi_p(x,t)=\braket{x}
{\phi_p(t)}$ with momentum $p$ for the electronic degree of freedom and plane waves $\chi_k(R,t)=
\braket{R}{\chi_k(t)}$ with momentum $k$ for the nuclear degree of freedom. The explicit forms of 
these wave functions, with normalizations $\delta(k-k')=\braket{\chi_k(t)}{\chi_{k'}(t)}$ and $\delta(p-p')=
\braket{\phi_p(t)}{\phi_{p'}(t)}$, are 
  \begin{align}
    \phi_p(x,t)&=(2\pi)^{-1/2} \exp{\left[ i \left( p x-\frac{p^2t}{2\mu}- \frac{p}{\mu}\int^tA(t')dt'\right)
    \right]},
    \label{eq:DI_sole}\\
    \chi_k(R,t)&=(2\pi)^{-1/2} \exp{\left[ i \left( kR-\frac{k^2t}{m_\text{p}} \right) \right]}.
    \label{eq:DI_solN}
 \end{align}

The wave packet $\ket{\Psi_\text{DI}(t)}$ is expanded in the direct product basis of Volkov and plane 
waves
\begin{equation}
  \begin{aligned}
    \ket{\Psi_\text{DI}(t)} = \theta_\text{e}\theta_\text{N}\ket{\Psi(t)}
    = \int dp \int dk  b_{p,k}(t) \ket{\phi_{p}(t)}\ket{\chi_{k}(t)},
  \end{aligned}
\end{equation}
where 
\begin{equation}
  b_{p,k}(T)= \bra{\phi_{p}(t)}\bra{\chi_{k}(t)} \theta_\text{e}\theta_\text{N} \ket{\Psi(t)}
\end{equation}
is the differential probability amplitude for measuring an electronic momentum $p$ and a nuclear 
momentum $k$. The joint momentum spectrum (JMS) reads
\begin{equation}
    \pdt{P}{p}{k}=\abs{b_{p,k}(T)}^2.
    \label{eq:DI-JMS}
\end{equation}
The corresponding JES that gives the differential probability for observing a nuclear KER of 
$E_\text{N}=k^2/m_\text{p}$ and an electron with energy $E_\text{e}=p^2/2\mu$ is given by
\begin{equation}
  \begin{aligned}
    \pdt{P}{E_\text{e}}{E_\text{N}}=\sum_{\text{sgn}(p)}\frac{m_\text{p}\mu}{2\abs{p}k}\abs{b_{p,k}(T)}^2,
    \label{eq:DI-JES}
  \end{aligned}
\end{equation}
where the summation over $\text{sgn}(p)$ refers to the summation of $\pm p$ corresponding to the 
same $E_\text{e}$.

The expression for $b_{p,k}(T)$ is rewritten using Eqs.~\eqref{eq:TDSE-DI}$-$\eqref{eq:TDSE-DI3} and 
the fundamental theorem of analysis. The result reads
\begin{equation}
  \begin{aligned}
    b_{p,k}(T)
    = b^\text{e}_{p,k}(T) + b^\text{N}_{p,k}(T)
    \label{eq:b_DI1}
  \end{aligned}
\end{equation}
with 
\begin{equation}
  \begin{aligned}
    b^\text{e}_{p,k}(T) = i \int_{-\infty}^T dt \bra{\phi_{p}(t)} \comm{T_\text{e}+V_\text{I}(t)}{\theta_
    \text{e}} \bra{\chi_{k}(t)} \theta_\text{N} \ket{\Psi(t)},
    \label{eq:b_DI1e}
  \end{aligned}
\end{equation}
and
\begin{equation}
  \begin{aligned}
    b^\text{N}_{p,k}(T) =i \int_{-\infty}^T dt \bra{\chi_{k}(t)} \comm{T_\text{N}}{\theta_\text{N}}
    \bra{\phi_{p}(t)}\theta_\text{e}\ket{\Psi(t)}.
    \label{eq:b_DI1N}
  \end{aligned}
\end{equation}
In Eqs.~\eqref{eq:b_DI1e} and \eqref{eq:b_DI1N}, the commutators vanish everywhere except at the 
discontinuity of the step functions. The probability amplitudes can thus be obtained by integrating the 
time-dependent surface flux. The amplitude $b_{p,k}^\text{e}(T)$ is the amplitude corresponding to the 
flux going from the dissociation region into the DI region (see Fig.~\ref{fig1}), while
$b_{p,k}^\text{N}(T)$ is the amplitude corresponding to the flux going from the ionization region into 
the DI region. The two amplitudes must be added coherently to obtain the total amplitude for DI. It is, 
however, possible to choose $R_\text{s}$ sufficiently large so that all the flux going into the DI region in 
Fig.~\ref{fig1} originate from the ionization region, and therefore we can set $b^\text{e}_{p,k}(T)=0$.
The commutators in Eqs.~\eqref{eq:b_DI1e} and \eqref{eq:b_DI1N} can be calculated explicitly, with
\begin{equation}
  \begin{aligned}
    \comm{T_\text{e}+V_\text{I}(t)}{\theta_\text{N}} 
    =&\int dx  \ket{x}\left[- \frac{1}{2\mu}\delta^{(1)}(\abs{x}-x_\text{s}) - \textrm{sgn}(x)
    \delta(\abs{x}-x_\text{s})\left(\frac{1}{\mu}\pd{}{x}+i\beta A(t)\right) \right] \bra{x},
    \label{eq:DI-ecomm}
  \end{aligned}
\end{equation}
and 
\begin{equation}
  \begin{aligned}
    \comm{T_\text{N}}{\theta_\text{N}} 
    =- \frac{1}{m_\text{p}}\int dR  \ket{R}\left[\delta^{(1)}(R-R_\text{s}) +2\delta(R-R_\text{s})
    \pd{}{R}\right] \bra{R},
    \label{eq:DI-Ncomm}
  \end{aligned}
\end{equation}
where $\delta$ and $\delta^{(1)}$ are the Dirac delta function and its first derivative, respectively.
After inserting Eq.~\eqref{eq:DI-Ncomm} into Eq.~\eqref{eq:b_DI1N}, evaluating the resulting integral 
and collecting terms in Eq.~\eqref{eq:b_DI1}, we obtain
\begin{equation}
  \begin{aligned}
    b_{p,k}(T)
    &=\frac{1}{m_\text{p}} \int_{-\infty}^T dt \chi_{k}^* (R_\text{s},t) \left[ k- i \pd{}{R} \right] 
    \bra{\phi_{p}(t)}\theta_\text{e}
    \ket{\Psi(t)}
    \at[\bigg]{R_\text{s}} .\\
    \label{eq:DI-b2}
  \end{aligned}
\end{equation}

To calculate the amplitude of Eq.~\eqref{eq:DI-b2}, the matrix element 
$\bra{\phi_{p}(t)} \theta_\text{e} \ket{\Psi(t)}$ must be 
evaluated at $R_\text{s}$. Direct projection of $ \theta_\text{e} \ket{\Psi(t)}$ on the Volkov waves 
$\ket{\chi_{p}(t)}$ is not an 
option as the electronic CAP [Eq.~\eqref{eq:CAP}] will absorb part of the wave function at 
$\abs{x}>x_\text{CAP}>x_\text{s}$. To 
circumvent this problem we expand $\bra{\phi_{p}(t)} \theta_\text{e} \ket{\Psi(t)}$ in an arbitrary time-
independent basis $\zeta_m(R)$,

\begin{equation}
  \begin{aligned}
     \bra{\phi_{p}(t)}\theta_\text{e}\ket{\Psi(t)}
    = \sum_m a_{p,m}(t) \ket{\zeta_m}
    \label{eq:DI-dexp}
  \end{aligned}
\end{equation}
with $a_{p,m}(t)=\bra{\zeta_m}\bra{\phi_{p}(t)}\theta_\text{e}\ket{\Psi(t)}$. In our calculations we use a 
sine basis for $\zeta_m$. By taking the time derivative of $a_{p,m}(t)$ and using 
Eq.~\eqref{eq:DI-ecomm} to evaluate the resulting commutator, it can be shown that $a_{p,m}(t)$ 
satisfies
\begin{equation}
  \begin{aligned}
    \d{}{t}{a}_{p,m}(t) = -i \sum_{m'}\bra{\zeta_m} T_\text{N}+V_\text{N} \ket{\zeta_{m'}} a_{p,m'}(t) +  
    f^{+}_{p,m}(t)+f^{-}_{p,m}(t)
    \label{eq:DI-ader}
  \end{aligned}
\end{equation}
with 
\begin{equation}
  \begin{aligned}
    f^{+}_{p,m}(t)=\phi_{p}^*(x_\text{s},t)\bigg[ \left(\frac{p}{2\mu}+\beta A(t)\right)-\frac{i}{2\mu}\pd{}{x} 
    \bigg]  \braket{\zeta_m}{\Psi(t)} \at[\bigg]{x_\text{s}}
    \label{eq:DI-fplus}
  \end{aligned}
\end{equation}
and
\begin{equation}
  \begin{aligned}
    f^{-}_{p,m}(t)=-\phi_{p}^*(-x_\text{s},t)\bigg[ \left(\frac{p}{2\mu}+\beta A(t)\right)-\frac{i}{2\mu}\pd{}{x} 
    \bigg]  \braket{\zeta_m}{\Psi(t)} \at[\bigg]{-x_\text{s}}.
    \label{eq:DI-fminus}
  \end{aligned}
\end{equation}
The terms $f^{+}_{p,m}(t)$ and $f^{-}_{p,m}(t)$ can be interpreted as flux terms, counting the flux going 
through the surfaces at $x=x_\text{s}$ and at $x=-x_\text{s}$, respectively. Equation \eqref{eq:DI-ader}
can be solved using any of the standard numerical techniques for solving ordinary differential
equations; in our calculations we use a fourth-order Runge-Kutta method with time step 0.05. 
The coefficients 
$a_{p,m}(t)$  give us information on the wave packet even in regions where the CAP is active 
($\abs{x}>x_\text{CAP}>x_\text{s}$), as seen in Eq.~\eqref{eq:DI-dexp}. When describing laser ionization, 
one of the two terms in $f_{p,m}(t)$ will usually be negligible and can therefore be ignored. For 
example if $p$ is positive, then  $f^{-}$ will be zero since in this case there is no incoming wave at
$x=-x_\text{s}$.

Inserting Eq.~\eqref{eq:DI-dexp} into Eq.~\eqref{eq:DI-b2}, we obtain the final expression for 
$b_{p,k}(T)$ determining the JMS and JES through Eqs.~\eqref{eq:DI-JMS} and \eqref{eq:DI-JES}:
\begin{equation}
  \begin{aligned}
    b_{p,k}(T)
    =\frac{1}{m_\text{p}} \sum_m \left[ k-i \pd{}{R}  \right] \zeta_m(R)\at[\bigg]{R_\text{s}}  \int_{-\infty}^T 
    \text{d}t  \chi_{k}^*(R_\text{s},t)a_{p,m}(t).
    \label{eq:DI-bfinal}
  \end{aligned}
\end{equation}

\subsection{t-SURFF for dissociation}
\label{sec:theoryD}
Consider now the dissociation process $\text{H}_2^+ \rightarrow \text{H} + p$.
The projected TDSE on the region describing dissociation without ionization reads (see Fig.~\ref{fig1})
\begin{equation}
  \begin{aligned}
    i \partial_t\ket{\Psi_\text{D}(t)}=H_\text{D}(t)\ket{\Psi(t)},
  \end{aligned}
\end{equation}
where we have defined the projected Hamiltonian $H_\text{D}(t)=(1-\theta_\text{e})\theta_\text{N}H(t)$. 
On the right-hand side $\ket{\Psi(t)}$ and not $\ket{\Psi_\text{D}(t)}$ appears 
[see Eq.~\eqref{eq:TDSE-DI}].

To obtain the dissociation-channel-specific nuclear KER spectrum we define the adiabatic 
BO basis states $\ket{\phi_{\text{el},i}}$ as the solutions to the electronic 
time-independent Schr\"{o}dinger equation with parametric dependence on $R$:
\begin{equation}
  \begin{aligned}
    \Big(T_\text{e}+V_\text{eN}+V_\text{N} \Big)\ket{\phi_{\text{el},i}}=E_{\text{el},i}(R)\ket{\phi_{\text{el},i}},
    \label{eq:elec_eq}
  \end{aligned}
\end{equation}
where $E_{\text{el},i}(R)$ is the $i$th electronic potential energy surface in the BO approximation. 
To ease notation we do not explicitly include the parametric dependence on $R$ in the BO states.

The wave packet $\Psi_\text{D}(x,R,t)$ is expanded in the BO basis as
\begin{equation}
  \begin{aligned}
    \ket{\Psi_\text{D}(t)}
    &=(1-\theta_\text{e})\theta_\text{N}\ket{\Psi(t)} \\
    &=\int dk \sum_{i}c_{i,k}(t)\ket{\chi_{k}(t)}\ket{\phi_{\text{el},i}(t)},
    \label{eq:psi_D2}
  \end{aligned}
\end{equation}
where $\ket{\chi_{k}(t)}$ is a plane wave with momentum $k$ given in Eq.~\eqref{eq:DI_solN}, $\ket{\phi_{\text{el},i}(t)}=\ket{\phi_{\text{el},i}}e^{-iE_{\text{el},i}(R)t}$, and 
\begin{equation}
  c_{i,k}(t)=\bra{\chi_{k}(t)}\bra{\phi_{\text{el},i}(t)} (1-\theta_\text{e})\theta_\text{N} \ket{\Psi(t)}.
\end{equation}
The bra-ket notation used here indicates integration with respect to both the electronic and nuclear 
degrees of freedom.

In Eq.~\eqref{eq:psi_D2}, all the trivial time dependence is included in $\ket{\phi_{\text{el},i}(t)}$ and 
$\ket{\chi_{k}(t)}$, while the non trivial time dependence due to the external field and flux going from 
the bound region into the dissociation region of Fig.~\ref{fig1} is included in the expansion coefficients 
$c_{i,k}(t)$. At time $T>T_\text{pulse}$, when all the dissociative parts of the wave packet have moved 
into the dissociative region, $c_{i,k}(T)$ describes the differential probability amplitude for the electron 
to be in the bound state $i$ and the nuclear degree of freedom to have momentum $k$.

The expression for $c_{i,k}(T)$ can be written as
\begin{equation}
  \begin{aligned}
    c_{i,k}(T)
    = c_{i,k}^\text{N}(T) + c_{i,k}^\text{e}(T) + c_{i,k}^\text{I}(T),
    \label{eq:D-c0}
  \end{aligned}
\end{equation}
with 
  \begin{align}
    c_{i,k}^\text{N}(T)&=i \int_{-\infty}^T dt \bra{\chi_{k}(t)} \comm{T_\text{N}}{\theta_\text{N}} 
    \bra{\phi_{\text{el},i}(t)} (1-\theta_\text{e}) \ket{\Psi(t)}, \label{eq:D-cN}\\
    c_{i,k}^\text{e}(T)&=i \int_{-\infty}^T dt  \bra{\phi_{\text{el},i}(t)} \comm{T_\text{e}+V_\text{eN}
    +V_\text{N}}{(1-\theta_\text{e})} \bra{\chi_{k}(t)} \theta_\text{N} \ket{\Psi(t)},  \label{eq:D-ce} \\
    c_{i,k}^\text{I}(T)&=-i \int_{-\infty}^T dt \bra{\phi_{\text{el},i}(t)}  \bra{\chi_{k}(t)} {(1-\theta_
    \text{e})} \theta_\text{N} V_\text{I}(t) \ket{\Psi(t)}. \label{eq:D-cI}
  \end{align}
In the derivation of Eq.~\eqref{eq:D-cN}, it is assumed that the action of the nuclear kinetic energy 
operator on the electronic BO states is neglected.
This is in accordance with the BO approximation wherein the first- and second-order derivatives of the 
electronic state with respect to $R$ are neglected.

In Fig.~\ref{fig1}, The amplitude $c_{i,k}^\text{N}(T)$ corresponds to the flux going from the bound 
region into the dissociation region  through the surface at $R=R_\text{s}$, while $c_{i,k}^\text{e}(T)$ 
corresponds to the flux going  through the surfaces 
at $x=\pm x_\text{s}$. The amplitude $c_{i,k}^\text{e}$ can thus be neglected if the dissociative wave 
packet never reaches the surface $x=\pm x_\text{s}$ at time $T$. 
In the pure dissociation process, the electron is localized near one of the protons, 
i.e., along the lines $x=\pm (1/2)R$. The previous condition can thus always be satisfied if we choose 
$x_\text{s}>R_\text{s}/2$.

The amplitude $c_{i,k}^\text{I}(T)$ in Eq.~\eqref{eq:D-cI} includes the time-dependent interaction 
$V_\text{I}(t)$. Let $T_\text{impact}$ be the instant at which the fastest part of the dissociative wave 
packet hits the surface $R=R_\text{s}$. Then $c_{i,k}^\text{I}(T)$ can be neglected as long as 
$T_\text{impact}\ge T_\text{pulse}$. This can be seen by rewriting Eq.~\eqref{eq:D-cI} as 
\begin{equation}
  \begin{aligned}
    c_{i,k}^\text{I}(T)
    &=-i \int_{-\infty}^T dt \biggl\{\bra{\phi_{\text{el},i}(t)}  \bra{\chi_{k}(t)} V_\text{I}(t) {(1-\theta_
    \text{e})} \theta_\text{N}  \ket{\Psi(t)} \biggr.\\
    &\qquad \qquad +\biggl. \bra{\phi_{\text{el},i}(t)}  \bra{\chi_{k}(t)} \comm{(1-\theta_\text{e})}{V_\text{I}
    (t)} \theta_\text{N}  \ket{\Psi(t)} \biggr\}\\
  \end{aligned}
  \label{eq:D-cint}
\end{equation}
with the commutator in the velocity gauge given by
\begin{equation}
  \begin{aligned}
    \comm{(1-\theta_\text{e})}{V_\text{I}(t)}=-i\beta A(t) \int dx \ket{x}  \textrm{sgn}(x)
    \delta(\abs{x}-x_\text{s}) \bra{x}.
  \end{aligned}
\end{equation}

For $T_\text{impact}\ge T_\text{pulse}$, both terms in Eq.~\eqref{eq:D-cint} are zero. The first term is 
zero because $\ket{\Psi_\text{D}(t)}=(1-\theta_\text{e})\theta_\text{N} \ket{\Psi(t)}=0$ for $t<T_
\text{pulse}$, while $V_\text{I}(t)=0$ for $t>T_\text{pulse}$. Similarly, the second term is zero because 
$\Psi(\pm x_\text{s},R,t)=0$. The condition $T_\text{impact}\ge T_\text{pulse}$ depends on the 
interaction $V_\text{I}(t)$ and can be satisfied by placing the $R_\text{s}$ appropriately. 

The final expression for $c_{i,k}(T)$ is then, using Eqs.~\eqref{eq:DI-Ncomm} and 
\eqref{eq:D-c0},
\begin{equation}
  \begin{aligned}
    c_{i,k}(T)
    &=\frac{1}{m_\text{p}} \int_{-\infty}^T dt \chi_{k}^* (R_\text{s},t) \left[ k- i\pd{}{R}  \right]\bra{\phi_{\text{el},i}(t)} (1-\theta_\text{e}) \ket{\Psi(t)} \at[\bigg]{R_\text{s}}.
    \label{eq:D-cfinal}
  \end{aligned}
\end{equation}
We see that the differential probability amplitude $c_{i,k}(T)$ can be calculated by monitoring the flux 
going through the surface $R=R_\text{s}$. Moreover, the electronic BO-states $\phi_{\text{el},i}(x;R)$ 
with parametric dependence on $R$ only has to be calculated at points close to $R_\text{s}$, reducing 
the numerical effort.

The BO states $\ket{\phi_{\text{el},i}},$ $i =0,1,...$ are even for $i$ even, and odd for $i$ odd. For 
sufficiently large $R$, the states become pairwise degenerate. It is therefore natural to order the states 
into pairs, each pair consisting of a gerade and an ungerade state, $\phi_{\text{el},i}^{s},$ $i=0,1,...$, with 
$s=g,u$. 
In the dissociation process, the
differential probability for the nuclei to have a nuclear KER $E_\textrm{N}$ and the
electron to be in the $i$th electronic state with parity $s$ is therefore
given by
\begin{equation}
  \begin{aligned}
    \pd{P_{i}^{s}}{E_\text{N}}=\frac{m_\text{p}}{2k} \abs{c_{i,k}^{s}}^2,
    \label{eq:D-NKE}
  \end{aligned}
\end{equation}
where the amplitude $c_{i,k}^{s}$ is given in Eq.~\eqref{eq:D-cfinal}.

\subsection{t-SURFF for dissociation --- two-surface model}
In numerous previous descriptions of the dissociation process of ${\text{H}_2}^+$, the BO approximation 
involving only two bound
electronic states has been used
~\cite{Jolicard92,Giusti-Suzor95,Peng05,Leth09,Thumm08,Niederhausen08,Niederhausen12}.   
By making the ansatz $\Psi(x,R,t)=G_0(R,t)\phi_{\text{el},0}(x;R)+G_1(R,t)\phi_{\text{el},1}(x;R)$ in 
Eq.~\eqref{eq:TDSE}, where $G_0(R,t)$ and $G_1(R,t)$ are the nuclear wave functions corresponding 
to the lowest gerade $\phi_{\text{el},0}(x;R)$ and ungerade $\phi_{\text{el},1}(x;R)$ electronic states, 
respectively, the TDSE becomes
\begin{equation}
  \begin{aligned} 
    i
    \begin{pmatrix}
      \partial_tG_0(R,t) \\
      \partial_tG_1(R,t)
    \end{pmatrix}
    =
    \begin{pmatrix}
      -\frac{1}{m_\text{p}}\pdd{}{R}+E_{\text{el},0}(R) &  U_\text{I}(t)\\
      U_\text{I}(t)                             & -\frac{1}{m_\text{p}}\pdd{}{R}+E_{\text{el},1}(R)
    \end{pmatrix}
    \begin{pmatrix}
      G_0(R,t) \\
      G_1(R,t)
    \end{pmatrix}.
    \label{eq:TDSE-twosurface}
  \end{aligned}
\end{equation}
We refer to this model as the two-surface model. Due to the neglect of the excited and continuum 
electronic states in the ansatz, the TDSE in Eq.~\eqref{eq:TDSE-twosurface} is not gauge invariant. 
The dynamics are only correctly described in the length gauge \cite{Giusti-Suzor95}, where $U_\text{I}
(t)=\beta_\textrm{LG}\bra{\phi_{\text{el},1}(R)}x\ket{\phi_{\text{el},0}(R)}F(t)$, with the electric field $F(t)=-
\partial_t A(t)$ and $\beta_\textrm{LG}=1+1/(2m_\text{p}+1)$.
Here, we propagate Eq.~\eqref{eq:TDSE-twosurface} using the split-operator FFT method of
Ref.~\cite{Schwendner97}. 

The KER spectrum in the two-surface model can also be obtained by the t-SURFF method. The dissociation-channel-specific differential probability amplitudes $c^\text{2-BO}_{i,k}(T)$, with $i=0,1$, corresponding to the two BO states, read
\begin{equation}
  \begin{aligned}
    c^\text{2-BO}_{i,k}(T) 
    &= \bra{\chi_k(T)}\theta_\text{N}\ket{G_i(T)}\\
    &=\frac{1}{m_\text{p}} \int_{-\infty}^T dt \chi_{k}^* (R_\text{s},t) \left[ k- i\pd{}{R}  \right]G_i(t) \at[\bigg]{R_\text{s}},
  \end{aligned}
\end{equation}
which is obtained using the techniques of the previous two sections.
As in Sec.~\ref{sec:theoryD}, we have assumed that the laser pulse is over at the time the dissociative 
wave packet first hits the surface $R_\text{s}$. The dissociative spectrum within the two-surface model 
obtained using t-SURFF was compared with the results in Ref.~\cite{Jolicard92} and a perfect match 
was observed.

\section{Results and Discussion}
\label{sec:results}
To demonstrate the t-SURFF method for ${\text{H}_2}^+$, the TDSE of Eq.~\eqref{eq:TDSE} is solved with 
${\text{H}_2}^+$ initially in the ground state. The ground state is obtained by propagation in imaginary time. 
Two different sets of laser pulse parameters are used: one set with $\lambda=400\; \text{nm}$, 
$N_\text{c}=10$, and $I=8.8\times 10 ^{13}$ W/cm$^2$; another with $\lambda=800\; \text{nm}$, 
$N_\text{c}=10$, and $I=8.8\times 10 ^{13}$ W/cm$^2$. For the first pulse the duration is 
$T_\text{pulse}=13.3$ fs, the photon energy is $\omega = 3.1$ eV and the Keldysh 
parameter~\cite{Keldysh64} is $\gamma=3.4$, indicating that the dynamics take place in the 
multiphoton ionization regime. For the second pulse the duration is $T_\text{pulse}=26.7$ fs, 
$\omega = 1.55$ eV, and $\gamma=1.7$, placing the dynamics closer to the tunneling regime. These 
pulse parameters are chosen to facilitate comparisons with recent calculations using the same model 
for ${\text{H}_2}^+$ \cite{Madsen12,Silva13}.
All the following results are obtained with $x_\text{s}=50$ and $R_\text{s}=25$. Convergence of all 
results are checked by performing calculations with varying grid spacings, simulation volumes, CAP 
parameters, placement of t-SURFF surfaces, and observing that the results match. The
convergence with respect to propagation time deserves special mentioning.
The time $T$, at which the dissociation and DI wave packets have moved inside their respective 
regions (see Fig.~\ref{fig1}), is written as $T=T_\textrm{pulse}+T_\textrm{free}$, with $T_\textrm{free}$ 
being the propagation time after the pulse. An estimate of $T_\textrm{free}$ for DI can be made by 
assuming the slowest nuclei to have $E_N=1/R_{{\textrm{cl},0}}$, where $R_{{\textrm{cl},0}}$ is the 
outer classical turning point of the ground vibrational state. An estimate for $T_\textrm{free}$ is then 
given by $R_\textrm{s}\sqrt{R_{{\textrm{cl},0}}m_\textrm{p}/2}=932$. To obtain a more accurate $T_
\textrm{free}$ numerical tests are performed, and it is found that choosing $T_\textrm{free}=1200$ will 
lead to the convergence of both the dissociation and the DI spectra.

\subsection{Results for Dissociative ionization}
\begin{figure}
  \centering
  \includegraphics{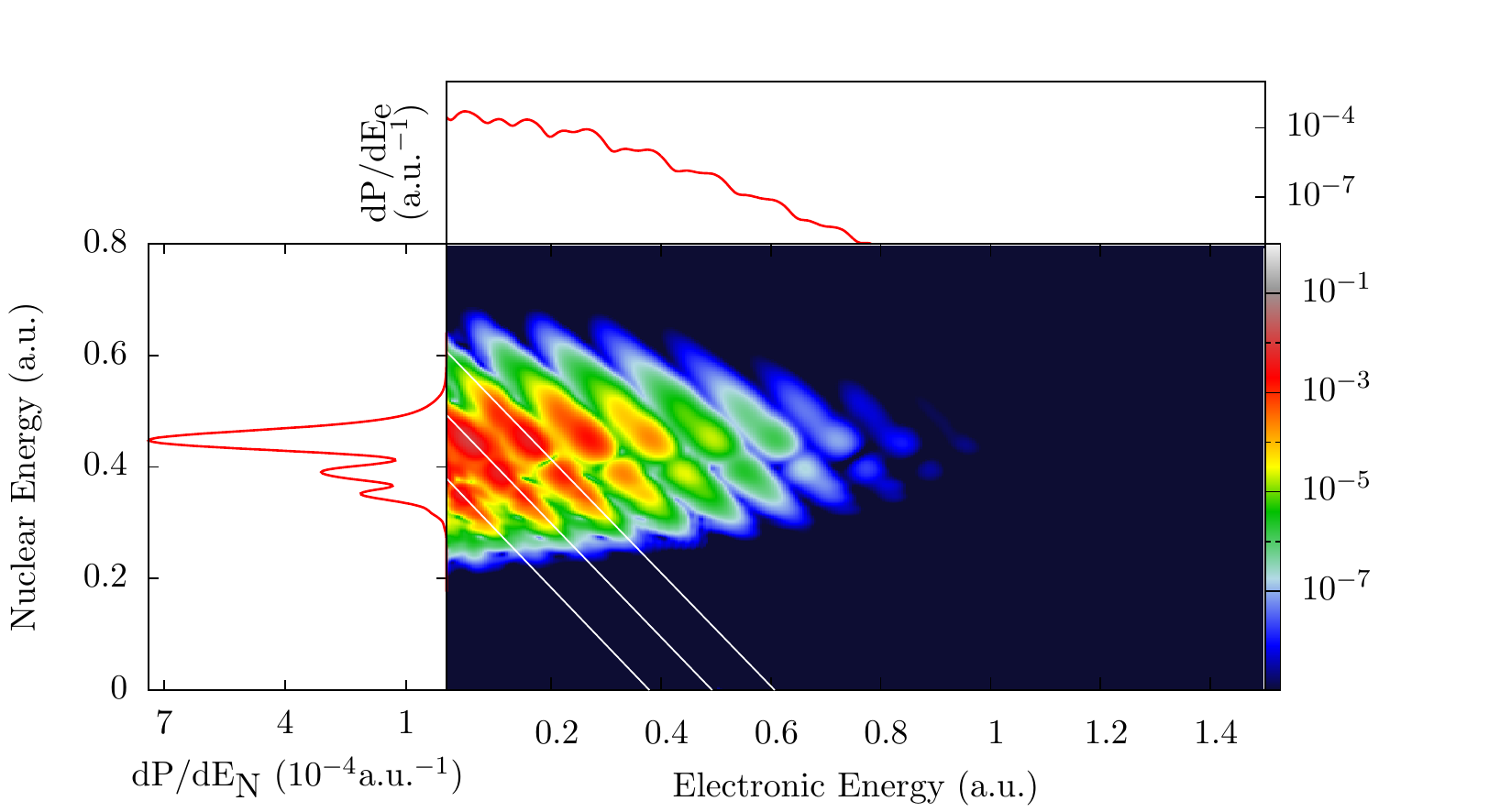}
  \caption{(Color online) JES for DI [Eq.~$\eqref{eq:DI-JES}$] for a pulse with parameters 
  $\lambda=400\; \text{nm}$, $N_\text{c}=10$, and $I=8.8\times 10 ^{13}$ W/cm$^2$. The top and left 
  side panels are the ATI and nuclear KER spectra, respectively. The diagonal (white) lines are energy 
  conservation lines  corresponding to $n$-photon absorption and satisfying $E_\text{N}+E_\text{e}
  =E_0+n\omega-U_\text{p}$, with the leftmost line corresponding to $n=9$.}
  \label{fig2}
\end{figure}

Figure \ref{fig2} shows the JES spectrum for the DI process of ${\text{H}_2}^+$  with laser parameters 
$\lambda=400\; \text{nm}$, 
$N_\text{c}=10$, and $I=8.8\times 10 ^{13}$ W/cm$^2$. We can compare Figure \ref{fig2} with the 
results presented in Fig.~1(a) of Ref.~\cite{Madsen12} with the same field parameters and a perfect 
match is observed. In Ref.~\cite{Madsen12}, an electric field $F(t)$ with a sine-squared envelope was 
used, whereas in the present work the vector potential $A(t)$ has a sine-squared envelope 
\eqref{eq:vecpot}. The agreement between the results is expected, as the large $N_\text{c}$ value 
makes the carrier envelope phase difference between the pulses insignificant. In Fig.~\ref{fig2} the 
energy conservation lines in the JES satisfying $E_\text{N}+E_\text{e}=E_0+n\omega-U_\text{p}$ are 
clearly seen, where $E_0=-0.597$ a.u. is the 
ground state energy of ${\text{H}_2}^+$ and $U_\text{p}=A_0^2/4=0.0483$ a.u. is the ponderomotive energy. 

\begin{figure}
  \centering
  \includegraphics{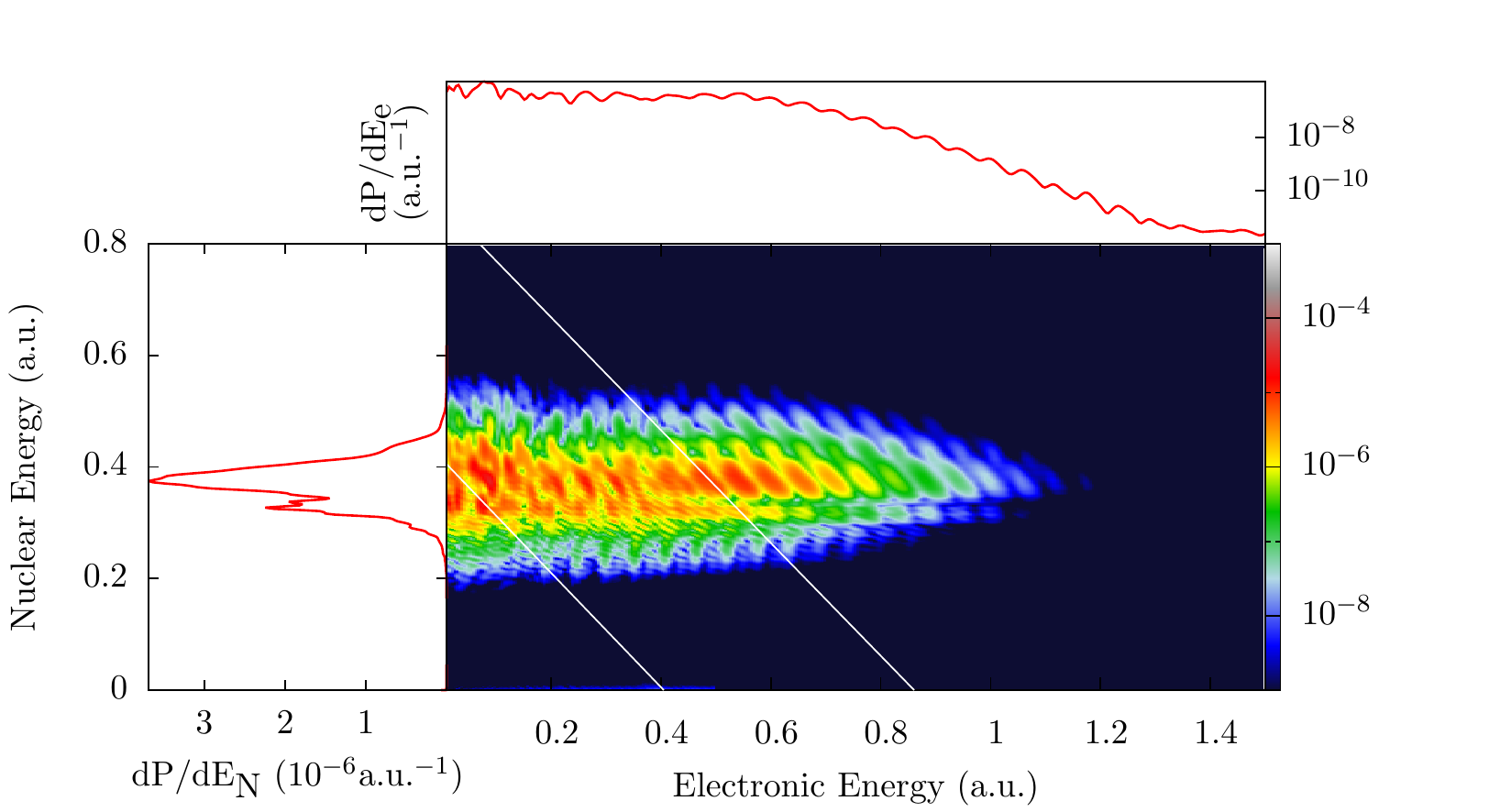}
  \caption{(Color online) Same as Fig.~\ref{fig2}, but for $\lambda=800\; \text{nm}$, $N_\text{c}=10$, 
  and $I=8.8\times 10 ^{13}$ W/cm$^2$.  The diagonal (white) lines are energy conservation lines 
  corresponding to $n$-photon absorption and satisfying 
  $E_\text{N}+E_\text{e}=E_0+n\omega-U_\text{p}$. The left diagonal (white) line corresponds to $n=21$, while the right diagonal (white) line corresponds to $n=29$.}
  \label{fig3}
\end{figure}

Figure \ref{fig3} shows the JES spectrum for the DI process for $\lambda=800\; \text{nm}$, 
$N_\text{c}=10$, and $I=8.8\times 10 ^{13}$ W/cm$^2$. For the part of the JES with 
$E_\text{e} \gtrsim 0.4$ a.u., the energy conservation lines are evident. At low electronic energies 
($E_\text{e} \lesssim 0.4$ a.u.), the energy conservation lines are not as clear. Instead, interference 
patterns are seen, corresponding to different pathways leading to the same final double-continuum 
state. Similar blurring of the photon-absorption lines was reported in Ref.~\cite{Silva13}, where the 
change in the shape of the spectrum was interpreted as a signature of tunneling ionization.
Figures~\ref{fig2} and \ref{fig3} show that the t-SURFF method can be used to describe DI.

\subsection{Results for dissociation}

\begin{figure}
  \centering
  \includegraphics{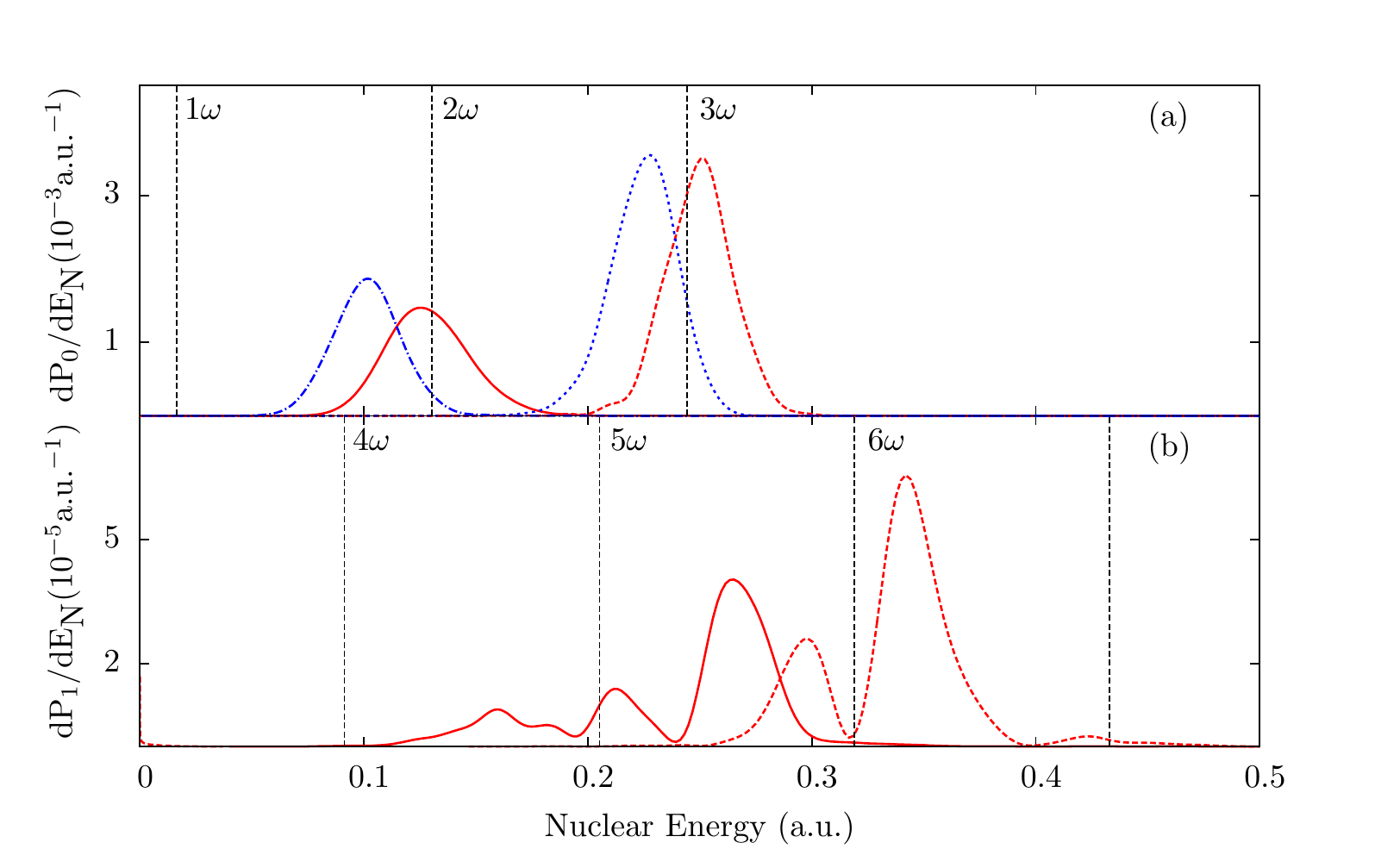}
  \caption{(Color online) Dissociation spectra [Eq.~\eqref{eq:D-NKE}] for a pulse with $\lambda=400\; 
  \text{nm}$, $N_\text{c}=10$, and $I=8.8\times 10 ^{13}$ W/cm$^2$ with dissociation via (a) the first 
  pair of gerade/ungerade states ($1s\sigma_g$ and $2p\sigma_u$), and (b) the second pair of gerade/
  ungerade states ($2s\sigma_g$ and $3p\sigma_u$). The solid (red) and the dashed (red) lines show 
  dissociation via the gerade and ungerade states, respectively, in the TDSE calculation. The dashed-
  dotted (blue) and the dotted (blue) lines show dissociation via the gerade and ungerade states, 
  respectively, in the two-surface BO calculation, scaled by a factor of 0.13. The vertical lines labeled 
  by $n\omega$ ($n=1,2,\dots,6$) denote photon absorptions above threshold (see text).}
  \label{fig4}
\end{figure}

\begin{figure}
  \centering
  \includegraphics{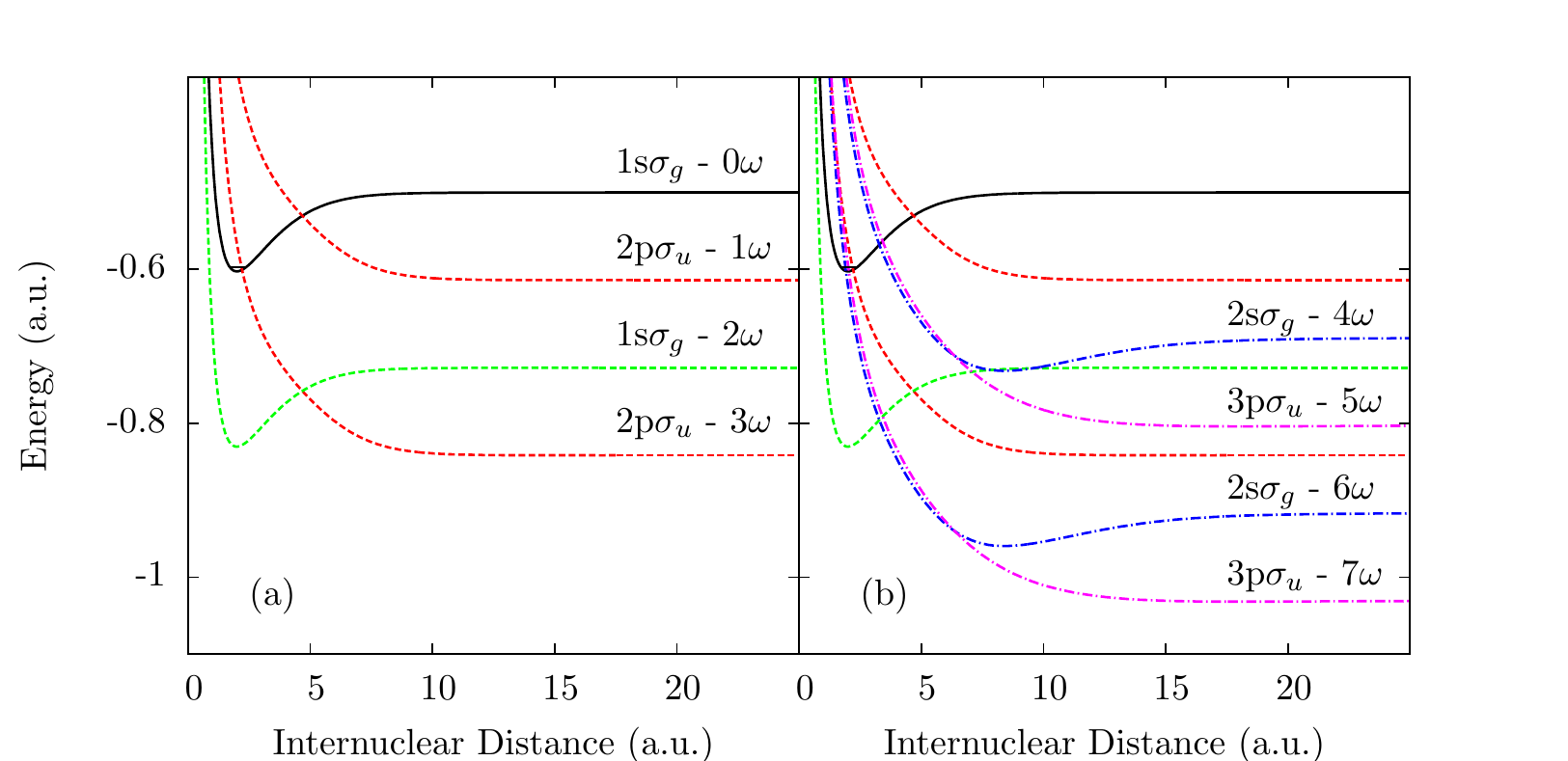}
  \caption{(Color online) The diabatic Floquet potentials for the one-dimensional ${\text{H}_2}^+$ model with $
  \lambda=400$nm. (a) Relevant dressed curves of $1s\sigma_g$ and $2p\sigma_u$ symmetry. (b) 
  Additional dressed curves of $2s\sigma_g$ and $3p\sigma_u$ symmetry.}
  \label{fig5}
\end{figure}

Figure \ref{fig4}(a) shows the nuclear KER spectrum for the 400 nm pulse with dissociation via the two 
first electronic states 
$\phi_{\text{el},0}^{g/u}$ corresponding to the $1s\sigma_g$ and $2p\sigma_u$ states.
A comparison of the magnitudes of the probabilities with Fig.~\ref{fig2} shows that dissociation 
dominates over the DI process for these field 
parameters, although the DI yield cannot be entirely neglected as done in the two-surface model. The 
vertical dashed lines in Fig.~\ref{fig4}(a) are the $n$-photon energy conservation lines satisfying 
$E_0+n\omega=E_{\text{el},0}(R=\infty)+E_\text{N}$, where $E_{\text{el},0}
(R=\infty)$ is the ground state energy of the hydrogen atom. Dissociation via $1s\sigma_g$ is located 
around the two-photon line, while dissociation via $2p\sigma_u$ is located around the three-photon line. This 
result can be understood by drawing the 
diabatic Floquet potential curves \cite{Giusti-Suzor95,Posthumus04}, shown in Fig.~\ref{fig5}(a).
Starting from the vibrational ground state, the laser can induce a dissociative wave packet by the ATD 
process, which will move down the $2p\sigma_u-3\omega$ curve.  The time for the wave packet to 
move from the intersection between $1s\sigma_g-0\omega$ and  $2p\sigma_u-3\omega$ at $R=2.23$ 
a.u. to the intersection between  $1s\sigma_g-2\omega$ and  $2p\sigma_u-3\omega$ at $R=4.7$ a.u. 
is approximately 204 a.u. (4.93 fs). At the latter intersection, part of the population can be transferred 
to the $1s\sigma_g-2\omega$ curve by stimulated photoemission. The time for the population to move 
from $R=4.7$ to $R=10$ via the surface $1s\sigma_g-2\omega$ is approximately 311 a.u. (7.52 fs), so 
the total time to reach $R=10$ a.u. from the vibrational ground state via the described pathway 
approximately equals the pulse duration. The pulse duration is therefore not long enough to induce 
transitions between the $1s\sigma_g-2\omega$ and  $2p\sigma_u-1\omega$ curves, nor is it intense 
enough to lower the adiabatic Floquet potentials (gap proportional to electric field) to induce tunneling 
from the vibrational ground state ($v=0$) to the $2p\sigma_u-1\omega$ curve. This is the reason for 
the absence of the one-photon peak in the nuclear KER spectrum.

In addition to the TDSE calculation, a calculation in the BO-approximation is performed for the 
two-surface model with the lowest pair of electronic states $1s\sigma_g$ and $2p\sigma_u$. In 
Fig.~\ref{fig4}(a) it is seen that the nuclear KER yield for this model is shifted more from the 
energy-conservation lines than the TDSE calculation, indicating that the ac-Stark shift is inaccurately 
accounted for in the two-surface model. Moreover, the dissociation yield is greatly overestimated by 
the two-surface model, which is understandable as in this model the excited electronic states together 
with the double continuum are completely neglected. The two-surface model is thus expected to be 
more accurate for lower laser intensities than for high intensities, where the coupling to the excited 
states and double continuum is strong.

Figure \ref{fig4}(b) shows the nuclear KER spectrum for the 400 nm pulse with dissociation via the 
third and fourth electronic states $\phi_{\text{el},1}^{g/u}$ corresponding to the $2s\sigma_g$ and 
$3p\sigma_u$ states. The vertical dashed lines in Fig.~\ref{fig4}(b) are the $n$-photon energy 
conservation lines satisfying $E_0+n\omega=E_{\text{el},1}(R=\infty)+E_\text{N}$, where 
$E_{\text{el},1}(R=\infty)$ is the energy of the first excited state in hydrogen. Dissociation via 
$2s\sigma_g$ is located between the four- and six-photon lines, while dissociation via 
$3p\sigma_u$ is located between the five- and seven-photon lines. 

\begin{figure}
  \centering
  \includegraphics{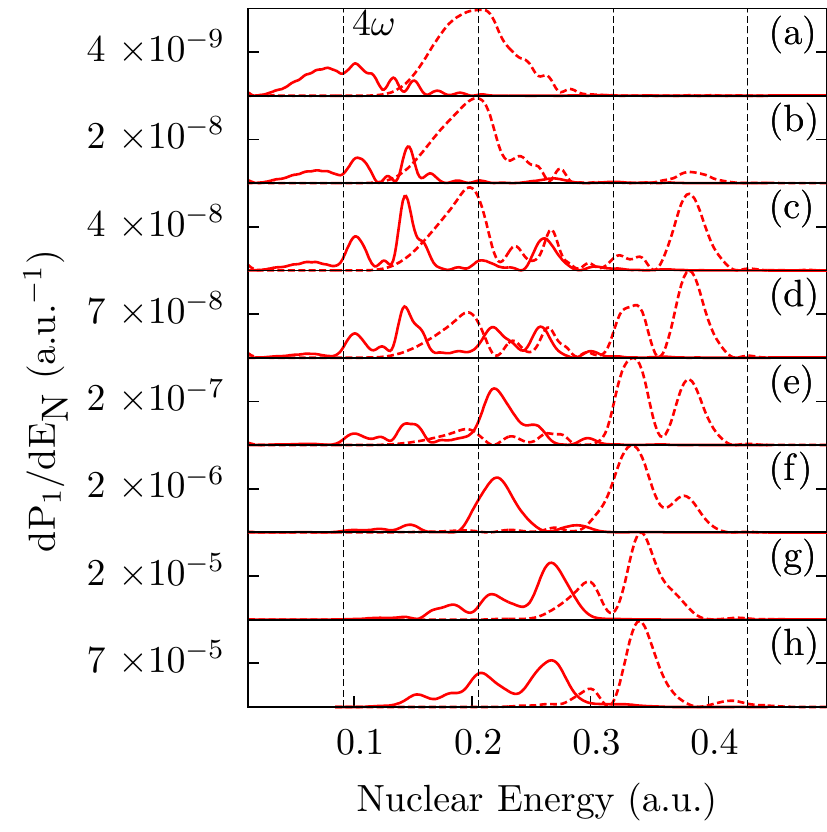}
  \caption{(Color online) Each subplot is as in Fig.~\ref{fig4}(b), now for the intensities: (a) $I=1\times 10 ^{13}$ W/cm$^2$,  (b) $I=2\times 10 ^{13}$ W/cm$^2$, (c) $I=3\times 10 ^{13}$ W/cm$^2$, (d) $I=3.5\times 10 ^{13}$ W/cm$^2$, (e) $I=4\times 10 ^{13}$ W/cm$^2$, (f) $I=5\times 10 ^{13}$ W/cm$^2$, (g) $I=8\times 10 ^{13}$ W/cm$^2$, and (h) $I=1\times 10 ^{14}$ W/cm$^2$.}
  \label{fig6}
\end{figure}

To understand the dissociation spectrum in Fig.~\ref{fig4}(b), the Floquet potential curves for $2s
\sigma_g$ and $3p\sigma_u$  dressed by four to seven photons are plotted in Fig.~\ref{fig5}(b).
Furthermore, a study is performed where we gradually increase the intensity of the laser field and 
observe the resulting nuclear KER spectra, shown in Fig.~\ref{fig6}. 
At lower intensities, $I \lesssim 2\times 10 ^{13}$~W/cm$^2$ in Figs. \ref{fig6}(a) and \ref{fig6}(b), the 
four-photon peak for the gerade state and the five-photon peak for the ungerade state are clearly seen, 
stemming from the wave packet following the pathway $1s\sigma_g-0\omega$ $\rightarrow$ $2p
\sigma_u-3\omega$  $\rightarrow$  $1\sigma_g-2\omega$ $\rightarrow$ $2\sigma_g-4\omega$ and  
$3p\sigma_u-5\omega$, shown in Fig. \ref{fig5}(b).
This is also indicated by the  $2s\sigma_g-4\omega$ and  $3p\sigma_u-5\omega$ curves in Fig.~
\ref{fig5}(b). In Fig.~\ref{fig6}(b) a small peak at around $E_\text{N}=0.4$ a.u. is seen for the ungerade 
state and a smaller peak at around $E_\text{N}=0.27$ a.u. is seen for the gerade state. These are the 
ac-Stark-shifted seven-photon and six-photon absorption peaks, respectively. Figure \ref{fig5}(b) clearly 
shows that the $3p\sigma_u-7\omega$ and $2s\sigma_g-6\omega$ curves cross the $1s
\sigma_g-0\omega$ curve at $R=2$ a.u., below the energy of the vibrational ground state ($v=0$), 
leading to 
 ATD processes explaining the peaks in Fig.~\ref{fig6}(b). As the intensity is increased from
Fig.~\ref{fig6}(c) to Fig.~\ref{fig6}(h), the seven-photon and eight-photon peaks are Stark shifted to lower 
nuclear energies, and additional structures in the peaks emerge. The additional structures are believed 
to be due to the interferences from the near degeneracy of the $3p\sigma_u-7\omega$ and $2s
\sigma_g-6\omega$ curves in Fig.~\ref{fig5}(b) for $R<7$ a.u., i.e., the strong coupling inducing many 
one-photon absorption and emission paths that all lead to the same final dissociating state. 

\begin{figure}
  \centering
  \includegraphics{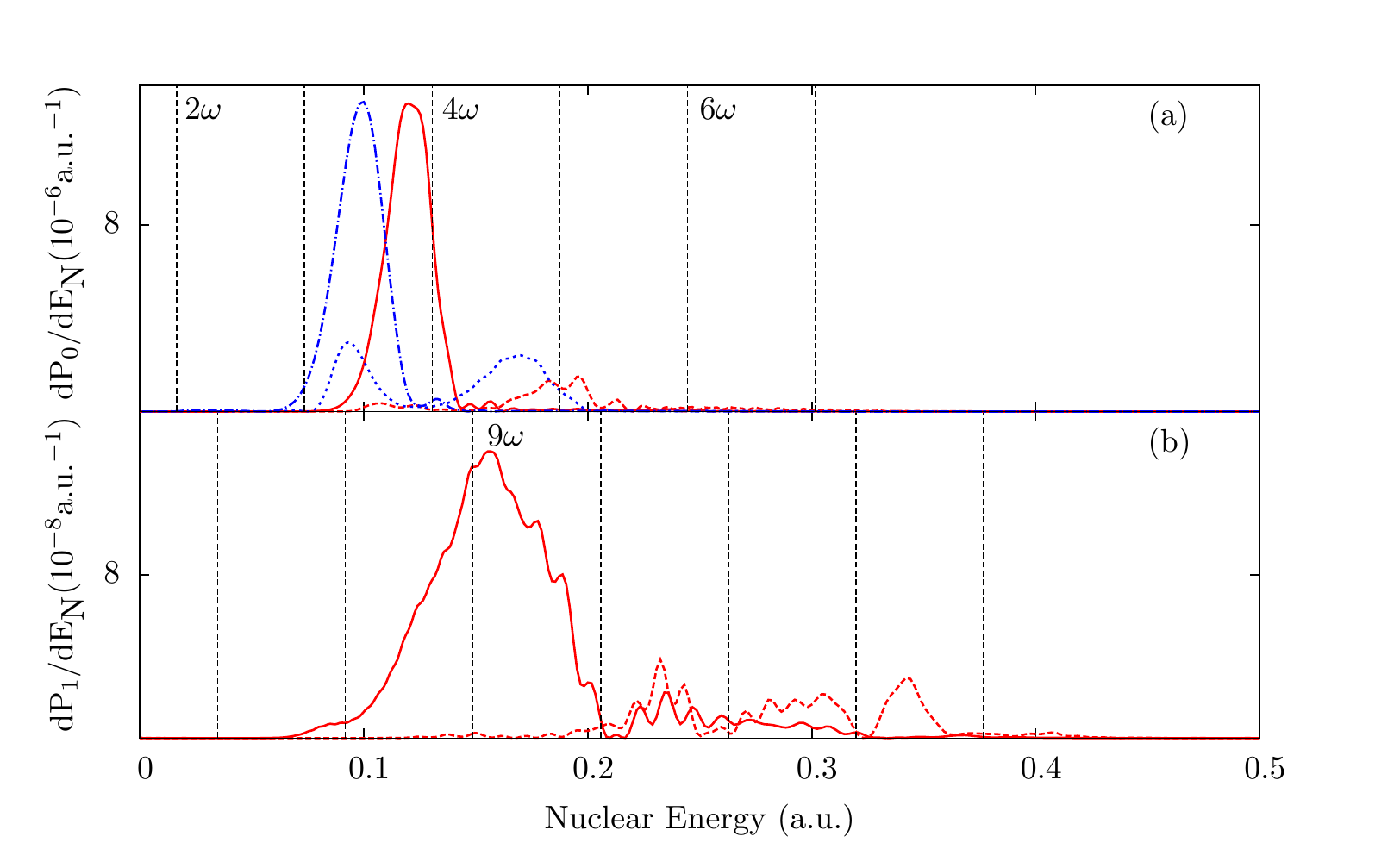}
  \caption{(Color online) Same as Fig.~\ref{fig4}, but for a pulse with parameters $\lambda=800\; \text{nm}$, $N_\text{c}=10$, and $I=8.8\times 10 ^{13}$ W/cm$^2$. }
  \label{fig7}
\end{figure}

\begin{figure}
  \centering
  \includegraphics{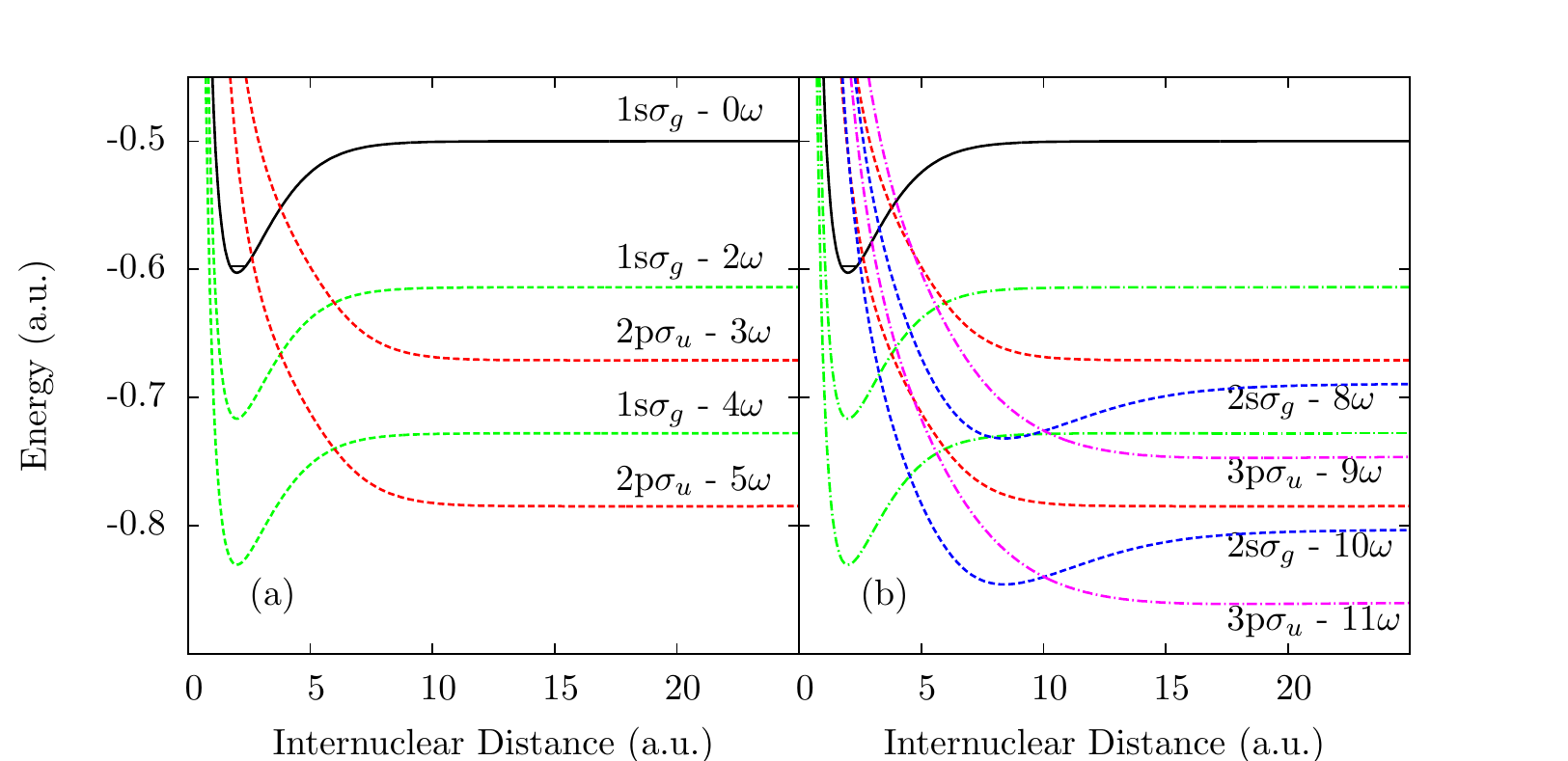}
  \caption{(Color online) Same as Fig.~\ref{fig5}, but for a pulse with parameters $\lambda=800\; \text{nm}$, $N_\text{c}=10$, and $I=8.8\times 10 ^{13}$ W/cm$^2$. }
  \label{fig8}
\end{figure}

Figure \ref{fig7}(a) shows the nuclear KER spectrum for a 10 cycle, 800 nm pulse with dissociation via the two first electronic states $\phi_{\text{el},0}^{g/u}$. Comparing with Fig.~\ref{fig3}, we see that the dissociation process is also the most dominant for these field parameters. Dissociation via $1s\sigma_g$ is located around the four-photon line, while dissociation via $2p\sigma_u$ is located around the five-photon line. This result can be understood by looking at the diabatic Floquet potential curves in Fig.~\ref{fig8}(a).

Starting from the vibrational ground state, coupling to the field can induce a dissociative wave packet 
on the $2p\sigma_u-5\omega$ surface, at a time when the intensity of the pulse is large enough for five-photon 
absorption. The time for the wave packet to move from the intersection between 
$1s\sigma_g-0\omega$ and $2p\sigma_u-5\omega$  at $R=2.61$ a.u. to the intersection between 
$1s\sigma_g-2\omega$ and  $2p\sigma_u-5\omega$ at $R=3.79$ a.u. is approximately 203 a.u. 
(4.919 fs). To reach the $1s\sigma_g-2\omega$ surface involves the emission of three photons, and the 
wave packet has therefore a larger probability of continuing along the dissociative 
$2p\sigma_u-5\omega$ surface until it hits the intersection between $2p\sigma_u-5\omega$ and 
$1s\sigma_g-4\omega$ at $R=6$ a.u. around 347 a.u. (8.38 fs). The wave packet then moves along 
the $2p\sigma_u-5\omega$ and $1s\sigma_g-4\omega$ to asymptotic distances and yields the four- and 
five-photon peaks in the nuclear KER spectrum of Fig.~\ref{fig7}.

Figure \ref{fig7}(b) shows the nuclear KER spectrum for the 800 nm pulse with dissociation via 
$2s \sigma_g$ and $3p\sigma_u$. Dissociation via $2s\sigma_g$ is seen to be located between the eight-photon 
and ten-photon lines, while dissociation via $3p\sigma_u$ is located between the ten-photon 
and 13-photon lines.
The $2s\sigma_g$ peak is due to dissociation after ten-photon absorption, while the  $3p\sigma_u$ 
peak is due to dissociation after 11-photon absorption. This is consistent with Fig.~\ref{fig8}(b), 
where the $3p\sigma_u-11\omega$ and $2s\sigma_g-10\omega$ crossings with  the 
$1s\sigma_g-0\omega$ curve are below the energy of the vibrational ground state ($v=0$).

\section{Conclusion}
\label{sec:conclusion}
We extended the t-SURFF method originally introduced for the determination of electronic spectra 
\cite{Tao12,Scrinzi12} to 
extract information from TDSE calculations for dissociation and DI of molecules. Using the example of 
${\text{H}_2}^+$, the JES of DI and the nuclear KER spectrum of the dissociation process were extracted and 
analyzed. 
Laser pulses with intensity 
$8.8\times 10^{13}$~W/cm$^2$, ten optical cycles, and wavelengths of $400$ and $800$ nm were 
used. For the shorter 
wavelength, the JES exhibited energy conservation lines described by $E_\text{N}+E_\text{e}=E_0+n
\omega-U_\text{p}$, 
while for the longer wavelength the conservation lines for electronic energies less than 0.4 a.u. were 
smeared out due to  interference effects. 
For the dissociation process, the nuclear KER spectra were explained qualitatively by looking at the 
diabatic Floquet potentials. It was found that the two-surface model, where only the first two electronic 
states in the BO approximation are 
taken into account, overestimates the ac-Stark shifts and the dissociation yields. In addition to 
dissociation via $1s\sigma_g$ 
and $2p\sigma_u$, the dissociation spectra via higher excited electronic states $2s\sigma_g$ and 
$3p\sigma_u$ were obtained. These spectra were explained qualitatively by observing the gradual 
change in the nuclear KER spectra as
the intensity was scanned from relatively low to high values.

The present work demonstrates that  the t-SURFF method is able to describe the break up of 
${\text{H}_2}^+$. 
The t-SURFF method requires modest spatial simulation volumes, and can be used to determine both 
photoelectron, KER 
and joint energy and momentum spectra.
The t-SURFF method requires small simulation volumes compared to other standard methods for the extraction of observables. For instance, in our present calculations using t-SURFF, a simulation volume of $\abs{x} \leq 100$ is used for the electronic coordinate, which is much smaller than  $|x| \leq 1500$ \cite{Madsen12} and  $|x| \leq 3000$  \cite{Silva13} used previously.
The simulation volume in the t-SURFF method can be further decreased by improving the complex absorber or implementing the infinite exterior complex scaling method \cite{Scrinzi10}.
 The relatively small simulation volume makes the method 
attractive for extension to 
molecules with more electronic and nuclear degrees of freedom.

\begin{acknowledgments}
This work was supported by the Danish Center for Scientific Computing, the Danish Natural Science Research Council (Grant No. 10-085430), and an ERC-StG (Project No. 277767 -- TDMET).
\end{acknowledgments}


%

\end{document}